\documentclass[a4paper,12pt]{article}

\usepackage{amsmath,amssymb,a4}
\usepackage{feynmf}

\allowdisplaybreaks[1]
\makeatletter

\def\section{\@startsection{section}{1}{\z@}{-3.25ex plus -1ex minus
             -.2ex}{1.5ex plus .2ex}{\normalfont\bfseries}}
\def\subsection{\@startsection{subsection}{1}{\z@}{-3.25ex plus -1ex
                minus -.2ex}{1.5ex plus .2ex}{\normalfont\itshape}}

\renewenvironment{thebibliography}[1]
         {\section*{References}\frenchspacing\small
          \begin{list}{[\arabic{enumi}]}
         {\usecounter{enumi}\parsep=2pt\topsep 0pt
         \settowidth{\labelwidth}{[#1]}
         \leftmargin=\labelwidth\advance\leftmargin\labelsep
         \rightmargin=0pt\itemsep=0pt\sloppy}}{\end{list}}
\makeatother

\renewcommand{\title}[1]{\null\vspace{10mm}\noindent
                         {\Large{\bf #1}}\vspace{8mm}}
\newcommand{\authors}[1]{\noindent{\large #1}\vspace{3mm}}
\newcommand{\address}[1]{{\center{\noindent\small\itshape #1\vspace{0mm}}}}

\begin{document}

\begin{titlepage}

\begin{flushright}
\begin{tabular}{l}
hep-th/0209253\\
 TUW-02-18
\end{tabular}
\end{flushright}

\begin{center}

\title{Space/time noncommutative field theories and causality}

\authors{H.~Bozkaya$^1$, P.~Fischer$^2$, H.~Grosse$^3$,
  M.~Pitschmann$^4$, V.~Putz$^5$, M.~Schweda$^6$,
  R. Wulkenhaar$^7$}

\address{$^{1,2,4,5,6}$Institut f\"ur
    Theoretische Physik, Technische Universit\"at Wien, \\ Wiedner
    Hauptstra\ss{}e  8--10, A-1040 Wien, Austria}

\address{$^3$Institut f\"ur
    Theoretische Physik, Universit\"at Wien, \\
    Boltzmanngasse 5,  A-1090 Wien, Austria}

\address{$^{5,7}$Max-Planck-Institut f\"ur 
Mathematik in den Naturwissenschaften, 
\\
Inselstra\ss{}e 22-26, D-04103 Leipzig, Germany}

\vspace{1cm}

\begin{abstract}
  
  As argued previously, amplitudes of quantum field theories on
  noncommutative space and time cannot be computed using na\"{\i}ve
  path integral Feynman rules. One of the proposals is to use the
  Gell-Mann--Low formula with time-ordering applied before performing
  the integrations. We point out that the previously given
  prescription should rather be regarded as an interaction point
  time-ordering. Causality is explicitly violated inside the region of
  interaction. It is nevertheless a consistent procedure, which seems
  to be related to the interaction picture of quantum mechanics. In
  this framework we compute the one-loop self-energy for a space/time
  noncommutative $\phi^4$ theory. Although in all intermediate steps
  only three-momenta play a r\^ole, the final result is manifestly
  Lorentz covariant and agrees with the na\"{\i}ve calculation.
  Deriving the Feynman rules for general graphs, we show, however,
  that such a picture holds for tadpole lines only.

\end{abstract}

\end{center}

\footnotetext[1]{work supported by
  ``Fonds zur F\"orderung der Wissenschaften'' (FWF) under contract P15463.}
\footnotetext[2]{fischer@hep.itp.tuwien.ac.at, work supported by
  ``Fonds zur F\"orderung der Wissenschaften'' (FWF) under contract P15463.} 
\footnotetext[3]{grosse@doppler.thp.univie.ac.at}
\footnotetext[4]{work supported by
  ``Fonds zur F\"orderung der Wissenschaften'' (FWF) under contract P15463.}
\footnotetext[5]{vputz@mis.mpg.de, work supported by ``Fonds zur
  F\"orderung der Wissenschaften'' (FWF) under contract P15015-TPH.}
\footnotetext[6]{mschweda@tph.tuwien.ac.at}
\footnotetext[7]{raimar.wulkenhaar@mis.mpg.de, Schloe\ss{}mann fellow. }

\end{titlepage}

\section{Introduction}

Quantum field theories on noncommutative spaces are full of
surprises, indicating that a true \emph{understanding} of quantum
field theory is still missing \cite{Wilson:1973jj}. This means, on the
other hand, that studying the quantisation of field theories on
noncommutative spaces we resolve the degeneracy of various methods
developed for commutative geometries: The outcomes of different
quantisation schemes ported to noncommutative geometries will no
longer coincide.  

At the moment we know of two major challenges. First, the evaluation
of Feynman graphs as a perturbative solution of the path integral
produces a completely new type of infrared-like singularities
\cite{Minwalla:1999px, Matusis:2000jf} in non-planar graphs. This can
be understood from the power-counting theorem \cite{Chepelev:2000hm}
for non-commutative (massive, Euclidian) field theories, which implies
the existence of two types (rings and commutants) of non-local
divergences.

Second, the case of a Minkowskian signature of the noncommutative
geometry (``space/time noncommutativity'') turns out to be involved.
It was pointed out in \cite{Bahns:2002vm} that in the Minkowskian
(non-degenerate) case the Wick rotation of Euclidian Green's function
does \emph{not} give a meaningful result, first of all because
unitarity would be lost \cite{Gomis:2000zz}. The reason is that the
Osterwalder-Schrader theorem \cite{Osterwalder:dx} does not hold.
Already in \cite{Doplicher:tu} there was given a proposal for a
correct quantisation of field theories on space/time noncommutative
geometries: Starting with interaction Hamiltonians on a Fock space
\begin{align}
H_I(t) = \int_{x^0=t} \!\! d^3 x \;:(\phi\star \phi \star \dots
\star \phi)(x):
\end{align}
(and averaging over the noncommutativity parameter) the contributions
to the scattering amplitudes were defines as the Dyson series
\begin{align}
G_n(x_1,\dots,x_k) :=
\frac{\mathrm{(-i)}^n}{n!}\int dt_1\dots dt_n \; 
\Big\langle 0\Big| T\phi(x_1)\dots\phi(x_k) H_I(t_1) \cdots H_I(t_n) 
\Big|0\Big\rangle\;,
\label{Dyson}
\end{align}
where $T$ denotes the time-ordering with respect to
$\{x_1^0,\dots,x_k^0,t_1,\dots,t_n\}$ and $|0\rangle$ the vacuum
state. Unitarity is preserved. In \cite{Bahns:2002vm} there was added
a second proposal, the iterative solution of the (interacting) field
equation (Yang-Feldman approach), which has the advantage of being
manifestly covariant. Unitarity is preserved as well. We refer, in particular,
to \cite{Rim:2002if}.

A third approach, the direct application of the Gell-Mann--Low formula for
Green's functions, 
\begin{align}
G_n(x_1,\dots,x_k) :=
\frac{\mathrm{i}^n}{n!} \int d^4z_1\dots d^4z_n \; 
\Big\langle 0\Big| T\phi(x_1)\dots\phi(x_k) \mathcal{L}_I(z_1) \cdots
\mathcal{L}_I(z_n) \Big|0\Big\rangle^{con}\;,
\label{GML}
\end{align}
where $\mathcal{L}_I$ is the interaction Lagrangian, was elaborated in
\cite{Liao:2002xc}. The superscript $^{con}$ means projection onto the
connected part. Unitarity was investigated in \cite{Liao:2002pj}.
That approach was called ``time-ordered perturbation theory'' in
\cite{Liao:2002xc}, a name which we find ambiguous. The time-ordering
in \cite{Liao:2002xc} is considered for external vertices and
\emph{interaction points} only, and not with respect to the \emph{actual
time-order of the fields} in the interaction Lagrangian.
We give in section~\ref{sec2} a few comments on
the two natural ways of time-ordering. The version used in
\cite{Liao:2002xc} is an \emph{interaction-point time-ordering}
(IPTO), it is explicitly acausal, and to distinguish from a true
\emph{causal time-ordering}.

Explicit calculations for the proposed quantisation schemes of space/time
noncommutative field theories are technically much more cumbersome than
Feynman graph computations. It is therefore desirable to extract a
powerful calculus out of the general schemes. In a first step one has
to familiarise oneself with the computational methods of the new
approach.

For that purpose we compute in this paper the one-loop two-point
function for a $\phi^4$ theory on noncommutative space-time. The
result of the indeed very lengthy but straightforward calculation in
interaction point time-ordered perturbation theory agrees with the
na\"{\i}ve path integral computation of the relevant Feynman
graph. Deriving in section~\ref{generalcase} the Feynman rules for
IPTO, we show, however, that this is true for tadpole lines only
(which should be removed anyway by normal ordering). 

We may speculate that taking the true causal time-ordering in the
Gell-Mann--Low formula one ends up with the usual Feynman rules
involving the causal Feynman propagator. It seems, therefore, that
causality and unitarity are mutually exclusive properties of space/time
noncommutative geometries.

\section{Comments on time-ordering and causality}
\label{sec2}

By ``noncommutative $\mathbb{R}^4$'' one understands the \emph{algebra}
$\mathbb{R}^4_\theta$ of Schwartz class functions on ordinary four-dimensional
space, equipped with the multiplication rule
\begin{align}
(f \star g)(x) &= \int d^4s \int \frac{d^4 l}{(2\pi)^4} \,
f(x -\tfrac{1}{2} \tilde{l}) \,g(x+s)\,\mathrm{e}^{\mathrm{i} ls}\;,
&  \tilde{l}^\nu := l_\mu \theta^{\mu\nu} \;.
\label{starprod}
\end{align}
The product (\ref{starprod}) characterised by a real skew-symmetric
translation-invariant tensor $\theta^{\mu\nu}=-\theta^{\nu\mu}$ of dimension
$[\text{length}]^2$ is associative and noncommutative, it is a
\emph{non-local} product on rapidly decreasing functions.

We consider a scalar field theory on $\mathbb{R}^4_\theta$ given by the
classical action
\begin{align}
\Sigma = \int d^4z \Big(\frac{1}{2} g^{\mu\nu} (\partial_\mu \phi \star 
\partial_\nu \phi)(z) - \frac{1}{2} m^2 (\phi \star \phi)(z)
+ \frac{g}{4!} (\phi \star \phi \star \phi \star \phi)(z)\Big)\;,
\label{class}
\end{align}
with $\phi \in \mathbb{R}^4_\theta$. By definition (\ref{starprod}) we have
\begin{align}
\big(\phi \star \phi \star  \phi \star \phi\big)(z) 
& = \int \prod_{i=1}^3 \Big( d^4 s_i \frac{d^4 l_i}{(2\pi)^4} 
\;\mathrm{e}^{\mathrm{i}l_i s_i}\Big)
\nonumber
\\*
& \qquad \times 
\phi(z{-}\tfrac{1}{2} \tilde{l}_1) 
\phi(z{+}s_1{-}\tfrac{1}{2} \tilde{l}_2) 
\phi(z{+}s_1{+}s_2{-}\tfrac{1}{2} \tilde{l}_3) 
\phi (z{+} s_1{+}s_2{+}s_3) \;.
\label{phi4}
\end{align}

If $g^{\mu\nu}$ is the Minkowskian metric
$g^{\mu\nu}=\mathrm{diag}(1,-1,-1,-1)$, we cannot simply Wick-rotate
the Euclidian Green's functions obtained by evaluation of the path
integral, see \cite{Bahns:2002vm}. Here we shall follow the proposal
of \cite{Liao:2002xc} and use the Gell-Mann--Low formula (\ref{GML})
to define the quantum field theory. However, one has to be more
careful with the definition of the time-ordering. Let us consider the
simplest case of the two-point function at first order in $g$,
\begin{equation}
G(x,y) = \frac{g}{4!} \int d^4z \,\Big\langle0\Big| 
T \big( \phi(x) \phi(y) \big(\phi
\star \phi \star  \phi \star \phi\big)(z)\big) \Big|0\Big\rangle \;.
\label{Gxy}
\end{equation}
(We put the missing factor $\mathrm{i}$ directly into the formula for
the element of the $S$-matrix.) In the same manner as on commutative
space-time, the integration over the interaction point is performed
\emph{after} taking the time-ordered product. Since the
$\star$-product for $\theta^{0i}\neq 0$ is non-local in time, one has
to say clearly what one understands under time-ordering. Let us
discuss this nuance for the geometrical situation relevant for
(\ref{Gxy}):
\begin{align}
\parbox{100mm}{\begin{picture}(100,50)
\put(3,3){\vector(1,0){90}}
\put(3,3){\vector(0,1){45}}
\put(-6,48){\mbox{\small time}}
\put(95,0){\mbox{\small space}}
\put(10,10){\mbox{\small$\times$}}
\put(13,7){\mbox{\small$\phi(z{+}s_1{+}s_2{-}\tfrac{1}{2} \tilde{l}_3)$}}
\put(70,36){\mbox{\small$\times$}}
\put(73,33){\mbox{\small$\phi(z{-}\tfrac{1}{2} \tilde{l}_1)$}}
\put(30,27){\mbox{\small$\times$}}
\put(33,24){\mbox{\small$\phi(z{+}s_1{-}\tfrac{1}{2} \tilde{l}_2)$}}
\put(20,40){\mbox{\small$\times$}}
\put(23,38){\mbox{\small$\phi(z{+}s_1{+}s_2{+}s_3)$}}
\put(15,20){\mbox{\small$\times$}}
\put(18,17){\mbox{\small$(\phi\star \phi\star\phi\star\phi)(z)$}}
\put(60,16){\mbox{\small$\times$}}
\put(63,13){\mbox{\small$\phi(y)$}}
\put(8,32){\mbox{\small$\times$}}
\put(11,29){\mbox{\small$\phi(x)$}}
  \end{picture}}
\label{pic}
\end{align}
This arrangement of fields corresponds to the following non-vanishing
contribution to the true time-ordering of (\ref{Gxy}):
\begin{align}
G_{(\ref{pic})}(x,y) &= 
\int d^4z  \int \prod_{i=1}^3 \Big( d^4 s_i \frac{d^4 l_i}{(2\pi)^4} 
\;\mathrm{e}^{\mathrm{i}l_i s_i}\Big)
\tau(s_1^0{+}s_2^0{+}s_3^0{+}\tfrac{1}{2}\tilde{l}_1^0)
\tau(z^0{-}\tfrac{1}{2}\tilde{l}_1^0{-}x^0)
\nonumber
\\*
& \qquad 
\times 
\tau(x^0{-}z^0{-}s_1^0{+}\tfrac{1}{2}\tilde{l}_2^0)
\tau(z^0{+}s_1^0{-}\tfrac{1}{2}\tilde{l}_2^0{-}y^0)
\tau(y^0{-}z^0{-}s_1^0{-}s_2^0{+}\tfrac{1}{2}\tilde{l}_3^0)
\nonumber
\\*
& \qquad 
\times\Big\langle0\Big| 
\phi (z{+} s_1{+}s_2{+}s_3) 
\phi(z{-}\tfrac{1}{2} \tilde{l}_1) \phi(x)
\phi(z{+}s_1{-}\tfrac{1}{2} \tilde{l}_2) 
\phi(y) 
\phi(z{+}s_1{+}s_2{-}\tfrac{1}{2} \tilde{l}_3) 
\Big|0\Big\rangle \;.
\label{GTxy}
\end{align}
Here, $\tau(t)$ denotes the step function $\tau(t)=1$ for $t> 0$ and
$\tau(t)=0$ for $t < 0$. There are $6!=720$ different contributions to
(\ref{Gxy}) when interpreting the time-ordering in the Gell-Mann--Low 
formula as the name suggests. The time-ordering guarantees that causal
processes only contribute to the $S$-matrix. Positive energy solutions
propagate forward in time and negative energy solutions backward.

There exists a modification of (\ref{Gxy}), where the time-ordering is
defined with respect to the \emph{interaction point}:
\begin{align}
G'_{(\ref{pic})}(x,y) &= 
\int d^4z  \int \prod_{i=1}^3 \Big( d^4 s_i \frac{d^4 l_i}{(2\pi)^4} 
\;\mathrm{e}^{\mathrm{i}l_i s_i}\Big)
\tau(x^0{-}z^0) \tau(z^0{-}y^0) 
\nonumber
\\*
& \qquad 
\times\Big\langle0\Big| 
\phi(x) \phi(z{-}\tfrac{1}{2} \tilde{l}_1) 
\phi(z{+}s_1{-}\tfrac{1}{2} \tilde{l}_2) 
\phi(z{+}s_1{+}s_2{-}\tfrac{1}{2} \tilde{l}_3) 
\phi (z{+} s_1{+}s_2{+}s_3) \phi(y)
\Big|0\Big\rangle \;.
\label{GTPxy}
\end{align}
There are now only $3!=6$ different contributions of this type. Since
the individual fields are now (in most of the cases) at the wrong
place with respect to the time-order, the interpretation (\ref{GTPxy})
of the Gell-Mann--Low formula violates causality. Now both energy
solutions propagate in any direction of time. There is, however, an
argument in favour of (\ref{GTPxy}): Contributions (\ref{Dyson}) to
the Dyson series are precisely ordered with respect to the time stamp
of the interaction Hamiltonians. It does not matter how the
time-dependence of the interaction Hamiltonian is produced from the
time-dependence of the constituents.

Since it is completely unclear how to \emph{derive} the Gell-Mann--Low
formula in the noncommutative setting, we have no guidance so far
whether (\ref{GTxy}) or (\ref{GTPxy}) (or none of the two) is the
correct one. The authors of \cite{Liao:2002xc} do not mention
(\ref{GTxy}). They use the exponential form of the $\star$-product,
which is a formal translation\footnote{The derivatives in the
  exponential form of the $\star$-product are generalised derivatives
  in the sense of distribution theory, not ordinary derivatives. As
  such one cannot apply the na\"{\i}ve rules of differential calculus.
  To make this transparent, write
  $\phi(x+a)\phi(y)=\exp(a^\mu\partial^x_\mu) \phi(x)\phi(y)$, and
  hide the exponential of the derivatives in the definition of the
  product.  It would be completely wrong to use the step function
  $\tau(x^0-y^0)$ or $\tau(y^0-x^0)$ for the product
  $\phi(x+a)\phi(y)$. One of the authors (R.W.) is grateful to Edwin
  Langmann for explaining this matter to him.} of a correct formula in
momentum space, but which might be dangerous in position space.  See
also the discussion in \cite{Wulkenhaar:2002ps}. Apart from avoiding
subtleties with generalised derivatives, the use of (\ref{phi4})
instead of the exponential form simplifies the calculations
considerably.

\section{The one-loop two-point function in ``interaction point time-ordered
  perturbation theory''}

Since the calculation of the sum of terms (\ref{GTPxy}) is (at least)
by a factor of 120 simpler than the calculation of the sum of terms
(\ref{GTxy}), we evaluate in this paper the one-loop two-point
function interpreted according to (\ref{GTPxy}). The name
``time-ordered perturbation theory'' used in \cite{Liao:2002xc} does
not seem appropriate to us, because the previous discussion shows that
this approach is precisely \emph{not} based on time-ordering. We
should better call it ``interaction point time-ordered perturbation
theory'', and use the symbol $T_I$ instead of the true causal
time-ordering $T$. The calculation can be shortened considerably when
starting directly from the Feynman rule (\ref{GammaE}) derived in
section~\ref{generalcase}. But without computing at least one example
one has little understanding for the starting point (\ref{EIV}) of the
general derivation.

With these remarks, the entire contribution to the one-loop two-point
function in noncommutative $\phi^4$ theory reads
\begin{align}
G(x,y) &= \frac{g}{4!} \int d^4z \,\Big\langle0\Big| 
T_I \big( \phi(x) \phi(y) \big(\phi
\star \phi \star  \phi \star \phi\big)(z)\big) \Big|0\Big\rangle 
\nonumber
\\
& =  \frac{g}{4!} \int d^4z \Big( \tau(x^0-y^0) \tau(y^0-z^0)  \,
\big\langle 0\big| \phi(x) \phi(y) \big(\phi \star \phi \star  
\phi \star \phi\big)(z) \big|0\big\rangle
\nonumber 
\\[-1ex]
&  \qquad\quad + \tau(x^0-z^0) \tau(z^0-y^0) \,\big\langle0\big| 
\phi(x) \big(\phi \star 
\phi \star  \phi \star \phi\big)(z)  \phi(y)\big|0\big\rangle 
\nonumber 
\\
& \qquad \quad+ \tau(y^0-x^0) \tau(x^0-z^0)  \,\big\langle0\big| \phi(y) \phi(x)  
\big(\phi \star \phi \star  \phi \star \phi\big)(z)\big|0\big\rangle 
\nonumber 
\\
& \qquad \quad+ \tau(y^0-z^0) \tau(z^0-x^0)\, \big\langle0\big| 
\phi(y) \big(\phi \star 
\phi \star  \phi \star \phi\big)(z) \phi(x) \big|0\big\rangle
\nonumber 
\\
& \qquad \quad+ \tau(z^0-x^0) \tau(x^0-y^0) \, \big\langle 0\big|
\big(\phi \star \phi
\star  \phi \star \phi\big)(z) \phi(x) \phi(y) \big| 0\big\rangle 
\nonumber 
\\*
& \qquad \quad + \tau(z^0-y^0) \tau(y^0-x^0) \, \big\langle0\big| 
\big(\phi \star \phi
\star  \phi \star \phi\big)(z) \phi(y) \phi(x) \big|0\big\rangle\Big)\;,
\label{start}
\end{align}
with the $\star$-product given in (\ref{phi4}).
We follow the usual strategy to obtain in the end the amputated
on-shell momentum-space one-loop two-point function.
We insert (\ref{phi4}) into (\ref{start}) and split each field (at given
position $x$) $\phi(x)=\phi^+(x)+\phi^-(x)$ into negative and positive
frequency parts, which have the property
\begin{align}
\phi^-(x)\big|0\big\rangle &=0\;, &
\big\langle 0\big| \phi^+(x) &=0\;.
\end{align}
Our conventions are listed in the Appendix, they are opposite to
\cite{Liao:2002xc}. It is convenient now to commute the $\phi^-$
to the right and the $\phi^+$ to the left, using the commutation rule
\begin{equation}    \label{commutator}
[\phi^-(x_1),\phi^+(x_2)] = D^+(x_1-x_2) \;,  
\end{equation}
where $D^+(x_1-x_2)$ is the positive frequency propagator
\begin{equation} \label{causalprop}
  D^+(x_1-x_2) =  \int \frac{d^3k}{(2\pi)^3 2 \omega_k} 
\;\mathrm{e}^{-\mathrm{i} k^+(x_1-x_2)}\;, 
\qquad \omega_k = \sqrt{\vec{k}^2+m^2}\;,
\end{equation}
and $k^+_{\mu}=(+\omega_k, -\vec{k})$ the positive energy on-shell
four-momentum. A lengthy but completely standard computation yields
\begin{align}
G(x,y) &= G^{con}(x,y) +  G^{discon}(x,y) \;,
\\
G^{discon}(x,y) 
& = \frac{g}{4!} \int d^4z \int \prod_{i=1}^3 \Big(d^4 s_i 
\frac{d^4 l_i}{(2\pi)^4} 
\;\mathrm{e}^{\mathrm{i}l_i s_i}\Big)
\Big\{ 
\Big(\tau(x^0-y^0) \tau(y^0-z^0) D^+(x{-}y) 
\nonumber
\\
&\qquad + \tau(x^0-z^0) \tau(z^0-y^0)  D^+(x{-}y) 
+ \tau(z^0-x^0) \tau(x^0-y^0)  D(x{-}y) \Big) 
\nonumber
\\
&
\quad + \big( x \leftrightarrow y \big) \Big\}
\Big( 
D^+({-}\tfrac{1}{2} \tilde{l}_2 {-}s_2{+}\tfrac{1}{2} \tilde{l}_3) 
D^+ ({-}\tfrac{1}{2} \tilde{l}_1{-}s_1{-}s_2{-}s_3) 
\nonumber
\\
& \qquad  \qquad 
+ D^+({-}\tfrac{1}{2} \tilde{l}_1{-}s_1{-}s_2
{+}\tfrac{1}{2} \tilde{l}_3) 
D^+ ({-}\tfrac{1}{2} \tilde{l}_2{-}s_2{-}s_3) 
\nonumber
\\
& \qquad \qquad
+ D^+({-}\tfrac{1}{2} \tilde{l}_1
{-}s_1{+}\tfrac{1}{2} \tilde{l}_2) 
D^+({-}\tfrac{1}{2} \tilde{l}_3 {-}s_3) \Big)\;,
\\
G^{con}(x,y) 
& = \frac{g}{4!} \int d^4 z \int \prod_{i=1}^3 
\Big(d^4 s_i \frac{d^4 l_i}{(2\pi)^4} 
\;\mathrm{e}^{\mathrm{i}l_i s_i}\Big)
\bigg\{\bigg(\tau(x^0-y^0) \tau(y^0-z^0)  
\nonumber
\\
& \quad \times \Big\{\Big(
D^+ ({-}\tfrac{1}{2} \tilde{l}_1 {-} s_1{-}s_2{-}s_3) 
D^+(x{-} z{-}s_1{+}\tfrac{1}{2} \tilde{l}_2) 
D^+(y{-}z{-}s_1{-}s_2{+}\tfrac{1}{2} \tilde{l}_3) 
\nonumber
\\
& \qquad + 
D^+({-}\tfrac{1}{2} \tilde{l}_1 {-}s_1{-}s_2
{+}\tfrac{1}{2} \tilde{l}_3) 
D^+(x{-} z{-}s_1{+}\tfrac{1}{2} \tilde{l}_2) 
D^+(y{-} z{-} s_1{-}s_2{-}s_3) 
\nonumber
\\
& \qquad + 
D^+({-}\tfrac{1}{2} \tilde{l}_1
{-}s_1{+}\tfrac{1}{2} \tilde{l}_2) 
D^+(x{-} z{-}s_1{-}s_2{+}\tfrac{1}{2} \tilde{l}_3) 
D^+(y{-} z{-} s_1{-}s_2{-}s_3) 
\nonumber
\\
& \qquad + 
D^+ ({-}\tfrac{1}{2} \tilde{l}_2 {-}s_2{-}s_3) 
D^+(x{-}z{+}\tfrac{1}{2} \tilde{l}_1) 
D^+(y{-} z{-}s_1{-}s_2{+}\tfrac{1}{2} \tilde{l}_3) 
\nonumber
\\
& \qquad + 
D^+({-}\tfrac{1}{2} \tilde{l}_2 {-}s_2{+}\tfrac{1}{2} \tilde{l}_3) 
D^+(x{-} z{+}\tfrac{1}{2} \tilde{l}_1) 
D^+(y {-} z{-} s_1{-}s_2{-}s_3) 
\nonumber
\\
& \qquad + 
D^+ ({-}\tfrac{1}{2} \tilde{l}_3 {-}s_3) 
D^+(x{-} z{+}\tfrac{1}{2} \tilde{l}_1) 
D^+(y{-}z{-}s_1{-}\tfrac{1}{2} \tilde{l}_2) \Big)
\nonumber
\\
& \quad +(x \leftrightarrow y) \Big\}
\nonumber
\\
& + \tau(x^0-z^0) \tau(z^0-y^0) 
\nonumber
\\
&\quad \times \Big\{ 
D^+ ({-}\tfrac{1}{2} \tilde{l}_1{-} s_1{-}s_2{-}s_3) 
D^+(x{-}z{-}s_1{-}s_2{+}\tfrac{1}{2} \tilde{l}_3) 
D^+(z{+}s_1{-}\tfrac{1}{2} \tilde{l}_2{-}y) 
\nonumber
\\
& \qquad + 
D^+({-}\tfrac{1}{2} \tilde{l}_2 {-}s_2{-}s_3) 
D^+(x{-} z{-}s_1{-}s_2{+}\tfrac{1}{2} \tilde{l}_3) 
D^+(z{-}\tfrac{1}{2} \tilde{l}_1{-}y)
\nonumber
\\
& \qquad + 
D^+({-}\tfrac{1}{2} \tilde{l}_1{-}s_1{-}s_2
{+}\tfrac{1}{2} \tilde{l}_3) 
D^+ (x{-}z{-} s_1{-}s_2{-}s_3) 
D^+(z{+}s_1{-}\tfrac{1}{2} \tilde{l}_2{-}y)
\nonumber
\\
& \qquad + 
D^+({-}\tfrac{1}{2} \tilde{l}_2 {-}s_2{+}\tfrac{1}{2} \tilde{l}_3) 
D^+ (x{-}z{-} s_1{-}s_2{-}s_3) 
D^+ (z{-}\tfrac{1}{2} \tilde{l}_1{-}y) 
\nonumber
\\
& \qquad + 
D^+({-}\tfrac{1}{2} \tilde{l}_1 {-} s_1{-}s_2{-}s_3) 
D^+ (x{-}z{-}s_1{+}\tfrac{1}{2} \tilde{l}_2) 
D^+ (z{+}s_1{+}s_2{-}\tfrac{1}{2} \tilde{l}_3{-}y)
\nonumber
\\
& \qquad + 
D^+({-}\tfrac{1}{2} \tilde{l}_3 {-}s_3) 
D^+ (x{-}z{-}s_1{+}\tfrac{1}{2} \tilde{l}_2) 
D^+ (z{-}\tfrac{1}{2} \tilde{l}_1{-}y)
\nonumber
\\
& \qquad  + 
D^+({-}\tfrac{1}{2} \tilde{l}_1
{-}s_1{+}\tfrac{1}{2} \tilde{l}_2) 
D^+ (x{-}z{-} s_1{-}s_2{-}s_3) 
D^+ (z{+}s_1{+}s_2{-}\tfrac{1}{2} \tilde{l}_3{-}y)
\nonumber
\\
& \qquad + 
D^+ ({-}\tfrac{1}{2} \tilde{l}_1
{-}s_1{-}s_2{+}\tfrac{1}{2} \tilde{l}_3) 
D^+(x{-}z{-}s_1{+}\tfrac{1}{2} \tilde{l}_2) 
D^+ (z{+} s_1{+}s_2{+}s_3{-}y) 
\nonumber
\\
& \qquad + 
D^+ ({-}\tfrac{1}{2} \tilde{l}_1 {-}s_1{+}\tfrac{1}{2} \tilde{l}_2) 
D^+(x{-}z{-}s_1{-}s_2{+}\tfrac{1}{2} \tilde{l}_3) 
D^+(z{+} s_1{+}s_2{+}s_3{-}y)
\nonumber
\\
& \qquad + 
D^+ ({-}\tfrac{1}{2} \tilde{l}_2 {-}s_2{-}s_3) 
D^+(x{-}z{+}\tfrac{1}{2} \tilde{l}_1) 
D^+(z{+}s_1{+}s_2{-}\tfrac{1}{2} \tilde{l}_3{-}y)
\nonumber
\\
& \qquad  + 
D^+ ({-}\tfrac{1}{2} \tilde{l}_3 {-}s_3) 
D^+(x{-}z{+}\tfrac{1}{2} \tilde{l}_1) 
D^+ (z{+}s_1{-}\tfrac{1}{2} \tilde{l}_2{-}y)
\nonumber
\\
& \qquad + 
D^+ ({-}\tfrac{1}{2} \tilde{l}_2 {-}s_2{+}
\tfrac{1}{2} \tilde{l}_3) 
D^+(x{-}z{+}\tfrac{1}{2} \tilde{l}_1) 
D^+ (z{+} s_1{+}s_2{+}s_3{-}y) \Big\} 
\nonumber
\\
& + \tau(z^0-x^0) \tau(x^0-y^0) 
\nonumber
\\
& \quad \times \Big\{ \Big(
D^+({-}\tfrac{1}{2} \tilde{l}_1 {-} s_1{-}s_2{-}s_3) \,
D^+(z{+}s_1{-}\tfrac{1}{2} \tilde{l}_2{-}x) 
D^+(z{+}s_1{+}s_2{-}\tfrac{1}{2} \tilde{l}_3{-}y) 
\nonumber
\\
& \qquad + 
D^+({-}\tfrac{1}{2} \tilde{l}_2 {-}s_2{-}s_3) \,
D^+(z{-}\tfrac{1}{2} \tilde{l}_1{-}x)
D^+(z{+}s_1{+}s_2{-}\tfrac{1}{2} \tilde{l}_3{-}y)
\nonumber
\\
& \qquad + 
D^+({-}s_3{-}\tilde{l}_3)\,
D^+(z{-}\tfrac{1}{2} \tilde{l}_1{-}x)
D^+(z{+}s_1{-}\tfrac{1}{2} \tilde{l}_2{-}y)
\nonumber
\\
& \qquad + 
D^+({-}\tfrac{1}{2} \tilde{l}_1 
{-}s_1{-}s_2{+}\tfrac{1}{2} \tilde{l}_3) 
D^+(z{+}s_1{-}\tfrac{1}{2} \tilde{l}_2{-}x) 
D^+(z{+} s_1{+}s_2{+}s_3{-}y)
\nonumber
\\
& \qquad + 
D^+({-}\tfrac{1}{2} \tilde{l}_2{-}s_2{+}\tfrac{1}{2} \tilde{l}_3) 
D^+(z{-}\tfrac{1}{2} \tilde{l}_1{-}x) 
D^+(z{+} s_1{+}s_2{+}s_3{-}y)
\nonumber
\\
& \qquad + 
D^+({-}\tfrac{1}{2} \tilde{l}_1
{-}s_1{+}\tfrac{1}{2} \tilde{l}_2) 
D^+(z{+}s_1{+}s_2{-}\tfrac{1}{2} \tilde{l}_3{-}x)
D^+(z{+} s_1{+}s_2{+}s_3{-}y)
\Big) 
\nonumber
\\
&\quad + \big(x \leftrightarrow y\big)\Big\}\bigg) 
+ \Big(x \leftrightarrow y\Big) \bigg\}\;.
\label{DDD}
\end{align}
We have to take the connected part $G^{con}(x,y)$ only. 
Inserting (\ref{causalprop}) we can perform the $s_i$-integrations, which
result in $\delta$-distributions in $l_i$, so that the $l_i$ integration can
be performed as well. The result has a remarkably compact form:
\begin{align}
G^{con}(x,y) 
& = \frac{g}{12} \int d^4z  
\int \frac{d^3 k_1}{(2\pi)^3 2\omega_{k_1}}
\int \frac{d^3 k_2}{(2\pi)^3 2 \omega_{k_2}}
\,\cos(\tfrac{1}{2} k_1^+ \tilde{k}_2^+) 
\nonumber
\\*
& \quad \times \Big( 
\tau(x^0-y^0) \tau(y^0-z^0) 
\mathrm{e}^{-\mathrm{i} k_1^+(x-z) }
\mathrm{e}^{-\mathrm{i} k_2^+(y-z) }
\mathcal{I}^{++}(k_1^+,k_2^+)
\nonumber
\\*
&\qquad + \tau(y^0-x^0) \tau(x^0-z^0) 
\mathrm{e}^{-\mathrm{i} k_1^+(x-z) }
\mathrm{e}^{-\mathrm{i} k_2^+(y-z) }
\mathcal{I}^{++}(k_1^+,k_2^+)
\nonumber
\\*
& \qquad + \tau(x^0-z^0) \tau(z^0-y^0)
\mathrm{e}^{-\mathrm{i} k_1^+(x-z) }
\mathrm{e}^{-\mathrm{i} k_2^+(z-y) }
\mathcal{I}^{+-}(k_1^+,k_2^+)
\nonumber
\\*
& \qquad + \tau(y^0-z^0) \tau(z^0-x^0)
\mathrm{e}^{-\mathrm{i} k_1^+(z-x) }
\mathrm{e}^{-\mathrm{i} k_2^+(y-z) }
\mathcal{I}^{-+}(k_1^+,k_2^+)
\nonumber
\\
& \qquad + \tau(z^0-x^0) \tau(x^0-y^0)
\mathrm{e}^{-\mathrm{i} k_1^+(z-x) }
\mathrm{e}^{-\mathrm{i} k_2^+(z-y) }
\mathcal{I}^{--}(k_1^+,k_2^+)
\nonumber
\\*
& \qquad + \tau(z^0-y^0) \tau(y^0-x^0)
\mathrm{e}^{-\mathrm{i} k_1^+(z-x) }
\mathrm{e}^{-\mathrm{i} k_2^+(z-y) }
\mathcal{I}^{--}(k_1^+,k_2^+)
\Big)\;,
\end{align}
where $(\tilde{k}^+)^\nu  \equiv (k^+)_\mu \theta^{\mu\nu}$ and 
\begin{align}
\mathcal{I}^{\kappa\lambda}(k_1^+,k_2^+) &= 
\int \frac{d^3 k}{(2\pi)^3 2 \omega_k}\,
\big( 3 
+ \mathrm{e}^{\mathrm{i}\kappa k_1^+\tilde{k}^+ 
+ \mathrm{i} \lambda k_2^+ \tilde{k}^+ }
+ \mathrm{e}^{\mathrm{i} \kappa k_1^+\tilde{k}^+ }
+ \mathrm{e}^{\mathrm{i} \lambda k_2^+\tilde{k}^+ }\big)\;,\qquad
\kappa,\lambda=\pm 1\;.
\label{Ikl}
\end{align}

Next we pass to the Fourier-transformed Green's function 
\begin{align}
G^{con}(p,q) &= \int d^4x \,d^4y \,
\mathrm{e}^{\mathrm{i} p x + \mathrm{i}q y}\, G^{con}(x,y)\;.   
\end{align}
We insert the identity (use the residue theorem) 
\begin{equation}    \label{Stepfunc}
\tau(x^0-y^0) = \lim_{\delta \to 0} \frac{\mathrm{i}}{2 \pi} 
\int_{-\infty}^{\infty} dt \frac{\mathrm{e}^{-\mathrm{i}t(x^0-y^0)}}{
t+\mathrm{i}\delta}
\end{equation}
and perform the integrations over $x,y,z$. The result is a host of 
$\delta$-distributions, which allow us to integrate over 
$\vec{k}_1,\vec{k}_2,t_1,t_2$:
\begin{align}
&G^{con}(p,q) 
\nonumber
\\*
&= \lim_{\delta_1,\delta_2 \to 0} 
\frac{g}{12} \Big(\frac{\mathrm{i}}{2\pi}\Big)^2 
\int d^4x\, d^4y\, d^4z \int_{-\infty}^\infty 
\frac{dt_1}{t_1+\mathrm{i}\delta_1}
\int_{-\infty}^\infty \frac{dt_2}{t_2+\mathrm{i}\delta_2}
\nonumber
\\*
& \times 
\int \frac{d^3 k_1}{(2\pi)^3 2 \omega_{k_1}}
\int \frac{d^3 k_2}{(2\pi)^3 2 \omega_{k_2}}
\,\cos(\tfrac{1}{2} k_1^+ \tilde{k}_2^+) 
\nonumber
\\
& \times \Big( 
\mathrm{e}^{\mathrm{i}\{ x^0(p_0-t_1-\omega_{k_1}) 
+ y^0(q_0+t_1-t_2-\omega_{k_2})
+ z^0(t_2+\omega_{k_1}+\omega_{k_2}) 
+ \vec{x}(\vec{k}_1-\vec{p})
+ \vec{y}(\vec{k}_2-\vec{q})
- \vec{z}(\vec{k}_1+\vec{k}_2)\}}
\mathcal{I}^{++}(k_1^+,k_2^+)
\nonumber
\\*
& \quad + \mathrm{e}^{\mathrm{i}\{ x^0(p_0+t_1-t_2-\omega_{k_1}) 
+ y^0(q_0-t_1-\omega_{k_2})
+ z^0(t_2+\omega_{k_1}+\omega_{k_2}) 
+ \vec{x}(\vec{k}_1-\vec{p})
+ \vec{y}(\vec{k}_2-\vec{q})
- \vec{z}(\vec{k}_1+\vec{k}_2)\}}
\mathcal{I}^{++}(k_1^+,k_2^+)
\nonumber
\\
& \quad + \mathrm{e}^{\mathrm{i}\{ x^0(p_0-t_1-\omega_{k_1}) 
+ y^0(q_0+t_2+\omega_{k_2})
+ z^0(t_1-t_2+\omega_{k_1} -\omega_{k_2}) 
+ \vec{x}(\vec{k}_1-\vec{p})
- \vec{y}(\vec{k}_2+\vec{q})
+ \vec{z}(\vec{k}_2-\vec{k}_1)\}}
\mathcal{I}^{+-}(k_1^+,k_2^+)
\nonumber 
\\
& \quad + \mathrm{e}^{\mathrm{i}\{ x^0(p_0+t_2+\omega_{k_1}) 
+ y^0(q_0-t_1-\omega_{k_2})
+ z^0(t_1-t_2-\omega_{k_1} +\omega_{k_2}) 
- \vec{x}(\vec{k}_1+\vec{p})
+ \vec{y}(\vec{k}_2-\vec{q})
+ \vec{z}(\vec{k}_1-\vec{k}_2)\}}
\mathcal{I}^{-+}(k_1^+,k_2^+)
\nonumber
\\
& \quad + \mathrm{e}^{\mathrm{i}\{ x^0(p_0+t_1-t_2+\omega_{k_1}) 
+ y^0(q_0+t_2+\omega_{k_2})
- z^0(t_1+\omega_{k_1} +\omega_{k_2}) 
- \vec{x}(\vec{k}_1+\vec{p})
- \vec{y}(\vec{k}_2+\vec{q})
+ \vec{z}(\vec{k}_1+\vec{k}_2)\}}
\mathcal{I}^{--}(k_1^+,k_2^+)
\nonumber
\\
& \quad + \mathrm{e}^{\mathrm{i}\{ x^0(p_0+t_2+\omega_{k_1}) 
+ y^0(q_0+t_1-t_2+\omega_{k_2})
- z^0(t_1+\omega_{k_1} +\omega_{k_2}) 
- \vec{x}(\vec{k}_1+\vec{p})
- \vec{y}(\vec{k}_2+\vec{q})
+ \vec{z}(\vec{k}_1+\vec{k}_2)\}}
\mathcal{I}^{--}(k_1^+,k_2^+)
\Big)
\nonumber
\\
&= \lim_{\delta_1,\delta_2 \to 0} 
\frac{g}{12} (2\pi)^4 \delta(p+q)
\nonumber
\\
& \times \Big( 
\frac{1}{p_0{-}\omega_p{+}\mathrm{i}\delta_1} \,
\frac{1}{\omega_p{+}\omega_q{-}\mathrm{i}\delta_2} 
\frac{\cos(\tfrac{1}{2} p^+ \tilde{q}^+)}{4 \omega_p \omega_q}
\mathcal{I}^{++}(p^+,q^+) 
\nonumber
\\
& \quad + 
\frac{1}{q_0{-}\omega_q{+}\mathrm{i}\delta_1} \,
\frac{1}{\omega_p{+}\omega_q{-}\mathrm{i}\delta_2} 
\frac{\cos(\tfrac{1}{2} p^+ \tilde{q}^+) }{4 \omega_p \omega_q}
\mathcal{I}^{++}(p^+,q^+) 
\nonumber
\\
& \quad + 
\frac{1}{p_0{-}\omega_p{+}\mathrm{i}\delta_1} \,
\frac{1}{q_0{+}\omega_q{-}\mathrm{i}\delta_2} 
\frac{\cos(\tfrac{1}{2} p^+ (-\tilde{q})^+) }{4 \omega_p \omega_q}
\mathcal{I}^{+-}(p^+,(-q)^+)
\nonumber 
\\
& \quad + 
\frac{1}{q_0{-}\omega_q{+}\mathrm{i}\delta_1} \,
\frac{1}{p_0{+}\omega_p{-}\mathrm{i}\delta_2} 
\frac{\cos(\tfrac{1}{2} (-p)^+ \tilde{q}^+) }{4 \omega_p \omega_q}
\mathcal{I}^{-+}((-p)^+,q^+)
\nonumber
\\
& \quad + 
\frac{1}{\omega_p {+}\omega_q{-}\mathrm{i}\delta_1} \,
\frac{1}{{-}q_0{-}\omega_q{+}\mathrm{i}\delta_2} 
\frac{\cos(\tfrac{1}{2} (-p)^+ (-\tilde{q})^+) }{4 \omega_p \omega_q}
\mathcal{I}^{--}((-p)^+,(-q)^+)
\nonumber
\\
& \quad + 
\frac{1}{\omega_p {+}\omega_q{-}\mathrm{i}\delta_1} \,
\frac{1}{{-}p_0{-}\omega_p{+}\mathrm{i}\delta_2} 
\frac{\cos(\tfrac{1}{2} (-p)^+ (-\tilde{q})^+) }{4 \omega_p \omega_q}
\mathcal{I}^{--}((-p)^+,(-q)^+)
\Big)\;.
\label{Gpq}
\end{align}
Note the appearance of $\delta(p{+}q)$ implementing conservation of
the four-momentum (translation invariance). We have used $\omega_{\pm
  k}=\omega_k$.

Following \cite{Liao:2002xc} we amputate the external legs by multiplying
(\ref{Gpq}) by the inverse propagators
$-\mathrm{i}(p_0^2-\omega_p^2)$ and $-\mathrm{i}(q_0^2-\omega_q^2)$. 
Using $(\pm k)^+= \pm k^\pm$, in particular the identity
\begin{align}
\mathcal{I}^{\pm\pm}((\pm p)^+,(\pm q)^+) =   
\int \frac{d^3 k}{(2\pi)^3 2 \omega_k}\,
\big( 3 
+ \mathrm{e}^{\mathrm{i} p^\pm \tilde{k}^+ 
+ \mathrm{i} q^\pm \tilde{k}^+ }
+ \mathrm{e}^{\mathrm{i} p^\pm \tilde{k}^+ }
+ \mathrm{e}^{\mathrm{i} q^\pm \tilde{k}^+ }\big) \equiv
\mathcal{I}(p^\pm,q^\pm) \:,
\label{Ipq}
\end{align}
we obtain
\begin{align}
(2\pi)^4 \delta(p+q) \Gamma(p,q) 
&=  -(p_0^2-\omega_p^2)(q_0^2-\omega_q^2) G(p,q)
\nonumber
\\
&= - \lim_{\delta_1,\delta_2 \to 0} 
\frac{g}{12} (2\pi)^4 \delta(p+q)\,(p_0^2-\omega_p^2)(q_0^2-\omega_q^2) 
\nonumber
\\
& \times \Big( 
\frac{1}{p_0{-}\omega_p{+}\mathrm{i}\delta_1} \,
\frac{1}{\omega_p{+}\omega_q{-}\mathrm{i}\delta_2} 
\frac{\cos(\tfrac{1}{2} p^+ \tilde{q}^+)}{4 \omega_p \omega_q}
\mathcal{I}(p^+,q^+) 
\nonumber
\\
& \quad + 
\frac{1}{q_0{-}\omega_q{+}\mathrm{i}\delta_1} \,
\frac{1}{\omega_p{+}\omega_q{-}\mathrm{i}\delta_2} 
\frac{\cos(\tfrac{1}{2} p^+ \tilde{q}^+) }{4 \omega_p \omega_q}
\mathcal{I}(p^+,q^+) 
\nonumber
\\
& \quad + 
\frac{1}{p_0{-}\omega_p{+}\mathrm{i}\delta_1} \,
\frac{1}{q_0{+}\omega_q{-}\mathrm{i}\delta_2} 
\frac{\cos(\tfrac{1}{2} p^+ \tilde{q}^-) }{4 \omega_p \omega_q}
\mathcal{I}(p^+,q^-)
\nonumber 
\\
& \quad + 
\frac{1}{q_0{-}\omega_q{+}\mathrm{i}\delta_1} \,
\frac{1}{p_0{+}\omega_p{-}\mathrm{i}\delta_2} 
\frac{\cos(\tfrac{1}{2} p^- \tilde{q}^+) }{4 \omega_p \omega_q}
\mathcal{I}(p^-,q^+)
\nonumber
\\
& \quad + 
\frac{1}{\omega_p {+}\omega_q{-}\mathrm{i}\delta_1} \,
\frac{1}{{-}q_0{-}\omega_q{+}\mathrm{i}\delta_2} 
\frac{\cos(\tfrac{1}{2} p^- \tilde{q}^-) }{4 \omega_p \omega_q}
\mathcal{I}(p^-,q^-)
\nonumber
\\
& \quad + 
\frac{1}{\omega_p {+}\omega_q{-}\mathrm{i}\delta_1} \,
\frac{1}{{-}p_0{-}\omega_p{+}\mathrm{i}\delta_2} 
\frac{\cos(\tfrac{1}{2} p^- \tilde{q}^-) }{4 \omega_p \omega_q}
\mathcal{I}(p^-,q^-)
\Big)\;.
\end{align}
Taking on-shell external momenta $p_0=\omega_p$ and $q_0=-\omega_q$
there survives a single term (the third one):
\begin{align}
\Gamma(p^+,q^-) =  
\lim_{p_0 \to \omega_p\,,\; q_0 \to -\omega_q} \Gamma(p,q)
&= \frac{g}{12} 
\cos(\tfrac{1}{2} p^+ \tilde{q}^-) \mathcal{I}(p^+,q^-)
\nonumber
\\*
&= \frac{g}{12} 
\int \frac{d^3k}{(2\pi)^3 2 \omega_k} \Big(4 + 2\cos (k^+ \tilde{p}^+)\Big)\;.
\label{on-shell}
\end{align}
In the last line we have used momentum conservation $p^+=-q^-$ and the
skew-symmetry of $\theta$. The remaining integral over $\vec{k}$
consists of a planar $\theta$-independent part and a non-planar
$\theta$-dependent part (the cosine). The planar part coincides (up to a
factor $\frac{2}{3}$) with the commutative result, it is divergent and
to be renormalised as usual by multiplicative renormalisation (or
better completely removed by normal ordering).

To compute the non-planar part, first note that
\begin{equation}
\cos(k^+\tilde{p}^+)= \cos\big(\omega_k \tilde{p}_0 
- \vec{k}\vec{\tilde{p}}\big) 
=   \cos\big(\omega_k \tilde{p}_0 \big)
\cos(\vec{k}\vec{\tilde{p}})
+ \sin\big(\omega_k \tilde{p}_0 \big)
\sin(\vec{k}\vec{\tilde{p}})\;,
\label{sin}
\end{equation}
where $\tilde{p}_0:= (\tilde{p}^+)_0$ and $\vec{\tilde{p}}=
\overrightarrow{\tilde{p}^+}$.  The uneven sine-term will drop under
the integral. Using the residue theorem we have 
\begin{align}
\frac{\mathrm{e}^{\mathrm{i} \omega_k \tilde{p}_0}}{2\omega_k}
&= \left\{\begin{array}{ll}
\displaystyle \lim_{\epsilon\to 0} \frac{1}{2\pi\mathrm{i}} 
\int_{-\infty}^\infty dk_0\; 
\frac{\mathrm{e}^{-\mathrm{i} k_0 \tilde{p}_0}}{
(k_0+\omega_k+\mathrm{i}\epsilon)
(k_0-\omega_k-\mathrm{i}\epsilon)}\quad   & \text{for }\tilde{p}_0 >0 \;,
\\[2ex]
\displaystyle \lim_{\epsilon\to 0} \frac{1}{2\pi\mathrm{i}} 
\int_{-\infty}^\infty dk_0\; 
\frac{-\mathrm{e}^{-\mathrm{i} k_0 \tilde{p}_0}}{
(k_0+\omega_k-\mathrm{i}\epsilon)
(k_0-\omega_k+\mathrm{i}\epsilon)}\quad   & \text{for } \tilde{p}_0 <0 \;,
\end{array}\right.
\label{NonFeyn}
\\*
\frac{\mathrm{e}^{-\mathrm{i} \omega_k \tilde{p}_0}}{2\omega_k}
&= \left\{ \begin{array}{ll}
\displaystyle \lim_{\epsilon\to 0} \frac{1}{2\pi\mathrm{i}} 
\int_{-\infty}^\infty dk_0\; 
\frac{-\mathrm{e}^{-\mathrm{i} k_0 \tilde{p}_0}}{
(k_0+\omega_k-\mathrm{i}\epsilon)
(k_0-\omega_k+\mathrm{i}\epsilon)}\quad & \text{for }\tilde{p}_0 >0
\;,
\\[2ex]
\displaystyle \lim_{\epsilon\to 0} \frac{1}{2\pi\mathrm{i}} 
\int_{-\infty}^\infty dk_0\; 
\frac{\mathrm{e}^{-\mathrm{i} k_0 \tilde{p}_0}}{
(k_0+\omega_k+\mathrm{i}\epsilon)
(k_0-\omega_k-\mathrm{i}\epsilon)}\quad   & \text{for }\tilde{p}_0 <0
\;.
\end{array}\right.
\label{Feyn}
\end{align}
Inserting (\ref{sin}), (\ref{NonFeyn}) and (\ref{Feyn}) into
(\ref{on-shell}) we obtain for the non-planar graph
\begin{align}
\Gamma_{non-planar}(p^+,q^-) &\equiv  
\frac{g}{6} 
\int \frac{d^3k}{(2\pi)^3 2 \omega_k} \cos (k^+ \tilde{p}^+)
\nonumber
\\
&= \lim_{\epsilon \to 0} 
\frac{g}{6} 
\int \frac{d^4k}{(2\pi)^4}\;\Re \Big(
\frac{\mathrm{i}}{k^2_0-(\vec{k}^2+m^2)+\mathrm{i}\epsilon}
\Big) \mathrm{e}^{-\mathrm{i} k \tilde{p}^+}\;,
\label{np}
\end{align}
independent of the sign of $\tilde{p}_0$.  The result (\ref{np}) can
obviously be obtained by Feynman rules, with the prescription that in
non-planar tadpoles the propagator to use is \emph{the real part of
  the Feynman propagator}. That real part is arithmetic mean of causal
and acausal propagators. The observed acausality is no surprise,
because according to (\ref{GTPxy}) the interaction time-ordering $T_I$
explicitly violates causality. As we shall see in
section~\ref{generalcase}, the just given Feynman rule is true for
tadpole lines only.

Apart from taking the real part, the evaluation of (\ref{np})
coincides with the computation in the ``na\"{\i}ve'' Feynman graph
approach.  Let us nevertheless repeat the steps. We employ
Zimmermann's $\epsilon$-trick
\begin{align}
\frac{1}{k^2-m^2+\mathrm{i}\epsilon} &\mapsto
\frac{1}{k_0^2+\omega_k^2(\mathrm{i}\epsilon{-}1)}
=\frac{\epsilon'{-}\mathrm{i}}{(\epsilon'{-}\mathrm{i})k_0^2
+\omega_k^2(\epsilon{-}\epsilon'{+}\mathrm{i}{+}
\mathrm{i}\epsilon \epsilon')}\;,
\end{align}
the denominator of which has for $\epsilon'<\epsilon$ a positive real
part, which allows us to introduce a Schwinger parameter:
\begin{align}
&\Gamma_{non-planar}(p^+,q^-)
\nonumber
\\
&= \Re\bigg(
\lim_{\epsilon \to 0, \epsilon'<\epsilon} \frac{\mathrm{i}g}{6}
\int \frac{d^4k}{(2\pi)^4}
\int_0^\infty \!\!\! d\alpha \,
(\epsilon'{-}\mathrm{i})\mathrm{e}^{-\alpha
\{(\epsilon'{-}\mathrm{i})k_0^2
+(\vec{k}^2+m^2)(\epsilon{-}\epsilon'{+}\mathrm{i}{+}
\mathrm{i}\epsilon \epsilon')\} -\mathrm{i} k_0 \tilde{p}_0 
+\mathrm{i} \vec{k} \vec{\tilde{p}}}\bigg)
\nonumber
\\
&=\Re\bigg(
\lim_{\epsilon \to 0, \epsilon'<\epsilon} 
\frac{\mathrm{i}g}{6(4\pi)^2} \frac{(\epsilon'{-}\mathrm{i})^{\frac{1}{2}}}{
(\epsilon{-}\epsilon'{+}\mathrm{i}{+}
\mathrm{i}\epsilon \epsilon')^{\frac{3}{2}}} 
\int_0^\infty \! \frac{d\alpha}{\alpha^2} \,\mathrm{e}^{-
\frac{\tilde{p}_0^2}{4\alpha (\epsilon'{-}\mathrm{i})}
- \frac{\vec{\tilde{p}}^2}{4\alpha (\epsilon{-}\epsilon'{+}\mathrm{i}{+}
\mathrm{i}\epsilon \epsilon')}
-\alpha m^2 (\epsilon{-}\epsilon'{+}\mathrm{i}{+}
\mathrm{i}\epsilon \epsilon')}\bigg)
\nonumber
\\
&=\Re\bigg(\lim_{\epsilon \to 0}
\frac{2\mathrm{i}g}{3(4\pi)^2} \frac{1}{(\mathrm{i}\epsilon{-}1)^{\frac{3}{2}}}
\sqrt{\frac{m^2 (\mathrm{i}\epsilon{-}1)}{
\tilde{p}_0^2+\frac{\vec{\tilde{p}}^2}{(\mathrm{i}\epsilon{-}1)}}}
\,
K_1\Big(\sqrt{m^2(\vec{\tilde{p}}^2+(\mathrm{i}\epsilon{-}1)\tilde{p}_0^2)}
\Big) \bigg)
\nonumber
\\
&= - \Re\Big(\frac{2g}{3(4\pi)^2}\sqrt{-\frac{m^2}{\tilde{p}^2}} 
K_1\big(\sqrt{-\tilde{p}^2m^2}\big)\Big)\;.
\label{UVIR}
\end{align}  
We have used $\int_0^\infty \frac{d\alpha}{\alpha^2}
\exp(-u\alpha-v/(4\alpha))=4\sqrt{(u/v)} K_1(\sqrt{uv})$ for $\Re u>0$
and $\Re v>0$. 

In the particular case where the external momentum $p$ is put
on-shell, we have
\begin{align}
-\tilde{p}^2 = \vec{\tilde{p}}^2-\tilde{p}_0^2 = 
(\theta_{i0}\sqrt{\vec{p}^2+m^2}  +  
\theta_{ij}p^j)^2-(\theta_{0j}p^j)^2 \geq 0\;,
\label{p=0}
\end{align}
because $\tilde{p}^\mu$ has to be space-like or null 
as a vector which is orthogonal to the time-like vector $p^\mu$. Thus, the
projection onto the real part in (\ref{UVIR}) is superfluous, and
(\ref{UVIR}) agrees exactly with the na\"{\i}ve Feynman rule
computation of the sum of graphs 
\begin{fmffile}{fmftime}
\begin{align}
\parbox{40mm}{\begin{picture}(40,20)
\put(0,0){\begin{fmfgraph}(40,20)
\fmfleft{l}
\fmfright{r}
\fmftop{t1,t2,t4,t3}
\fmfbottom{b1,b2,b4,b3}
\fmf{plain,tension=2}{r,i}
\fmf{plain,tension=1}{i,l}
\fmffreeze
\fmf{plain,right=0.5}{i,t2}
\fmf{plain,right=0.7,rubout=5}{t2,b2}
\fmf{plain,right=0.5}{b2,i}
\end{fmfgraph}}
\put(1,12){\mbox{\small$p$}}
\put(25,17){\mbox{\small$k$}}
\end{picture}}
\quad + \quad
\parbox{40mm}{\begin{picture}(40,20)
\put(0,0){\begin{fmfgraph}(40,20)
\fmfleft{l}
\fmfright{r}
\fmftop{t1,t2,t4,t3}
\fmfbottom{b1,b2,b4,b3}
\fmf{plain,tension=2}{r,i}
\fmf{phantom,tension=1}{i,l}
\fmffreeze
\fmf{plain,right=0.5}{i,t2}
\fmf{plain,right=0.7}{t2,b2}
\fmf{plain,right=0.5}{b2,i}
\fmf{plain,rubout=5}{i,l}
\end{fmfgraph}}
\put(1,12){\mbox{\small$p$}}
\put(25,17){\mbox{\small$k$}}
\end{picture}}
\quad.
\end{align}
\end{fmffile}%
However, if these graphs appear as subgraphs in a bigger graph, the
momentum $p$ will be the off-shell momentum through a propagator, and
the projection to the real part makes a difference. 

\section{The general case}
\label{generalcase}

The graph we have computed (for off-shell external momenta!) is very
often made responsible for the so-called UV/IR mixing. In fact the
situation is more complex, as it is very well described in
\cite{Chepelev:2000hm}. The ultimate goal must be to derive the
power-counting theorem for interaction point time-ordered perturbation
theory (for noncommutative space and time). In a first step one has to
derive graphical rules to assign an integral to a given graph. 

Let us therefore consider the momentum integral for a general Feynman graph
for a noncommutative $\phi^4$ theory. 
A given connected contribution to the $E$-point function 
at order $V$ in the coupling constant has after performing the Wick
contractions, insertion of the $D^+$ according to (\ref{causalprop}),
integration over $s_i$ and $l_i$ appearing in (\ref{phi4}) and
insertion of step functions (\ref{Stepfunc}) the form
\begin{align}
G(x_1,\dots,x_E) &= \lim_{\epsilon\to 0}
\int \prod_{v=1}^V \frac{g\,d^4z_v}{4!}
\int \prod_{e=1}^E \frac{d^3 p_e}{(2\pi)^3 2\omega_{p_e}}
\int \prod_{i=1}^I \frac{d^3 k_i}{(2\pi)^3 2\omega_{k_i}}
\int \prod_{s=1}^{E+V-1} \frac{\mathrm{i}\,d t_s}{
(2\pi)(t_s{+}\mathrm{i}\epsilon)}
\nonumber
\\*
& \times 
\exp\Big(-\mathrm{i} \sum_{v=1}^V \sum_{s=1}^{E+V-1} T_{vs} z_v^0 t_s -
\mathrm{i} \sum_{e=1}^E \sum_{s=1}^{E+V-1} T_{es} x_e^0 t_s \Big)
\nonumber
\\*
& \times 
\exp\Big(-\mathrm{i} \sum_{v=1}^V z_v \Big(\sum_{i=1}^I J_{vi} k_i^+ 
+\sum_{e=1}^E J_{ve} p_e^+\Big)\Big)
\exp\Big( -\mathrm{i} \sum_{e=1}^E \sigma_e p_e^+ x_e\Big)
\nonumber
\\*
& \times 
\exp\Big(\mathrm{i} \theta^{\mu\nu} \Big(
\sum_{i,j=1}^I I_{ij} k_{i,\mu}^+ k_{j,\nu}^+ 
+ \sum_{i=1}^I \sum_{e=1}^E I_{ie} k_{i,\mu}^+ p_{e,\nu}^+ 
+ \sum_{e,f=1}^E I_{ef} p_{e,\mu}^+
p_{f,\nu}^+\Big)\Big)\;.
\label{EIV}
\end{align}
There are $E{+}V{-}1$ step functions according to the time differences
of the $E$ external points $x_e$ and the $V$ interaction points $z_v$.
For each $s$ there are two non-vanishing $T_{*s}$, where these two
indices $*$ are either two indices $e$, one index $e$ and one index
$v$, or two indices $v$. The $T_{*s}$ for which the vertex $*$ ($z_v$
or $x_e$) is later equals $+1$, the other one $-1$.  This gives the
second line in (\ref{EIV}). An external point $x_e$ is linked via the
external line with momentum $p_e$ to exactly one vertex $z_v$, i.e.\ 
for given $e$ there is a single non-vanishing $J_{ve}$. For our
$\phi^4$ theory there are $I=2V-\frac{1}{2}E$ internal lines ($E$ is
even) with momentum $k_i$ which link a vertex $z_v$ to another vertex
$z_{v'}$.  Thus, if $v\neq v'$ (no tadpoles) for given $i$ there are
two non-vanishing $J_{vi}$, whereas for $v=v'$ we have $J_{vi}k_i^+
\equiv 0$.  We orient the internal and external lines forward in time.
Then, the incidence matrices $J_{vi}, J_{ve}$ equal $-1$ if the line
leaves $v$ and $+1$ if the line arrives at $v$. Similarly,
$\sigma_e=-1$ if the line $e$ leaves $x_e$ and $\sigma_e=+1$ if the
line $e$ arrives at $x_e$.  The matrices $I_{ij}, I_{ie}, I_{ef}$ are
the intersection matrices \cite{Filk:dm, Chepelev:2000hm}, which
instead of the Euclidian rosette construction are in IPTO obtained as
follows: According to the definition (\ref{phi4}) of the
$\star$-product, write at each vertex $v$ the four fields in
(\ref{phi4}) as a time-sequence where $z_v{-}\frac{1}{2}\tilde{l}_1$
is the latest point and $z_v{+}s_1{+}s_2{+}s_3$ the earliest
point\footnote{By the way, this defines the time-orientation of
  tadpole lines.}, irrespective of the \emph{actual time-order} of
these four points.  Connect these points with vertices $y_1,y_2,y_3,v_4$
according to the following picture:
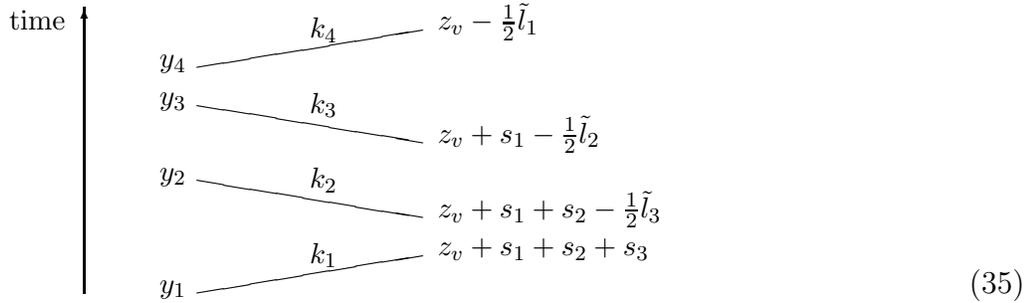
\begin{align}
\begin{picture}(100,40)  
\put(20,0){\line(6,1){30}}
\put(20,15){\line(6,-1){30}}
\put(20,25){\line(6,-1){30}}
\put(20,30){\line(6,1){30}}
\put(5,0){\vector(0,1){38}}
\put(-5,35){\mbox{\small time}}
\put(15,0){\mbox{\small$y_1$}}
\put(15,15){\mbox{\small$y_2$}}
\put(15,25){\mbox{\small$y_3$}}
\put(15,30){\mbox{\small$y_4$}}
\put(52,5){\mbox{\small$z_v+s_1+s_2+s_3$}}
\put(52,10){\mbox{\small$z_v +s_1+s_2-\frac{1}{2}\tilde{l}_3$}}
\put(52,20){\mbox{\small$z_v +s_1-\frac{1}{2}\tilde{l}_2$}}
\put(52,35){\mbox{\small$z_v -\frac{1}{2}\tilde{l}_1$}}
\put(35,4){\mbox{\small$k_1$}}
\put(35,14){\mbox{\small$k_2$}}
\put(35,24){\mbox{\small$k_3$}}
\put(35,34){\mbox{\small$k_4$}}
\end{picture}
\end{align}
The phase factor produced by the $s_n$ and $l_n$ variables is then
given by
\begin{align}
&\int \prod_{n=1}^3 \Big( d^4s_n \frac{d^4 l_n}{(2\pi)^4}
\;\exp(\mathrm{i} s_n l_n)\Big)
\nonumber
\\
&\quad \times \exp\Big(-\mathrm{i} k_1^+(s_1{+}s_2{+}s_3)J_{v1}   
-\mathrm{i} k_2^+(s_1{+}s_2{-}\tfrac{1}{2}\tilde{l}_3)J_{v2}   
-\mathrm{i} k_3^+(s_1{-}\tfrac{1}{2}\tilde{l}_2)J_{v3}   
-\mathrm{i} k_4^+({-}\tfrac{1}{2}\tilde{l}_1)J_{v4}  \Big)
\nonumber
\\
&= \exp\Big(\frac{\mathrm{i}}{2} \theta^{\mu\nu} \sum_{j=2}^4 
\sum_{i=1}^{j-1} k_{i,\mu}^+ J_{vi} k_{j,\nu}^+ J_{vj}\Big)
\equiv \exp\Big(\frac{\mathrm{i}}{2} \theta^{\mu\nu} \sum_{i,j=1}^4 
\tau^v_{ij}
k_{i,\mu}^+ J_{vi} k_{j,\nu}^+ J_{vj}\Big)\;. 
\end{align}
We have to define $\tau^v_{ij}=+1$ if the line $i$ is connected to an
``earlier'' field $\phi$ in the vertex $v$ than the line
$j$, otherwise $\tau^v_{ij}=0$. Summing over all vertices and
distinguishing external and internal lines, we are led to the
following identification in (\ref{EIV}):
\begin{align}
I_{ij}&=\frac{1}{2} \sum_{v=1}^V \tau^v_{ij} J_{vi}J_{vj}\;, &   
I_{ie}&= \frac{1}{2} \sum_{v=1}^V \big(\tau^v_{ie}-\tau^v_{ei}\big) 
J_{vi}J_{ve}\;, &   
I_{ef}&=\frac{1}{2} \sum_{v=1}^V \tau^v_{ef} J_{ve}J_{vf}\;.
\label{rosette}
\end{align}
Once more we notice the enormous computational advantage of using the
$\star$-product in the form (\ref{starprod}). 

We perform the Fourier transformation $\int \prod_{e=1}^E \big(d^4x_e\,
\exp(\mathrm{i}q_e x_e)\big)$ of (\ref{EIV}) to external momentum
variables $q$ as well as the $z_v$ integrations:
\begin{align}
  G(q_1,\dots,q_E) &= \lim_{\epsilon \to 0} \frac{g^V}{(4!)^V}
\prod_{e=1}^E \frac{1}{2\omega_{q_e}}
\int \prod_{i=1}^I \frac{d^3 k_i}{(2\pi)^3 2\omega_{k_i}}
\int \prod_{s=1}^{E+V-1} \frac{\mathrm{i}\,d t_s}{
(2\pi)(t_s+\mathrm{i}\epsilon)}
\nonumber
\\*
& \times 
\prod_{v=1}^V (2\pi)^3 \delta^3 \Big(\sum_{i=1}^I J_{vi} \vec{k}_i 
+\sum_{e=1}^E J_{ve}\sigma_e \vec{q}_e\Big)
\prod_{e=1}^E (2\pi) \delta \Big(q^0_e - \sigma_e \omega_{q_e}
- \sum_{s=1}^{E+V-1} T_{es} t_s \Big)
\nonumber
\\*
& \times 
\prod_{v=1}^V (2\pi)\delta \Big(\sum_{i=1}^I J_{vi} \omega_{k_i} 
+\sum_{e=1}^E J_{ve} \omega_{q_e} + \sum_{s=1}^{E+V-1} T_{vs} t_s \Big)
\nonumber
\\*
& \times 
\exp\Big(\mathrm{i} \theta^{\mu\nu} \Big(
\sum_{i,j=1}^I I_{ij} k_{i,\mu}^+ k_{j,\nu}^+ 
+\! \sum_{i=1}^I \sum_{e=1}^E 
I_{ie} \sigma_e k_{i,\mu}^+ q_{e,\nu}^{\sigma_e} 
+ \! \sum_{e,f=1}^E I_{ef} \sigma_e \sigma_f q_{e,\mu}^{\sigma_e}
q_{f,\nu}^{\sigma_f}\Big)\Big)\,.
\label{GE}
\end{align}
The vectors $\vec{q}_e$ are always outgoing from internal vertices.
There are now $E{+}V$ time-component $\delta$-functions involving the
$E{+}V{-}1$ integration variables $t_s$, after integration over which
there is one remaining $\delta$-function for the energy conservation
$\delta(\sum_{e=1}^E q_e^0)$. We multiply (\ref{GE}) by the inverse
propagators $\prod_{e=1}^E (-\mathrm{i})(q_e^2-\omega_{q_e}^2)$, remove 
$(2\pi)^4\delta^4(\sum_{e=1}^E q_e)$ by convention and put
$q_e^0=\sigma_e\omega_{q_e}$.  There is a non-vanishing contribution
only if the external vertices $x_e$ are either before or after the
internal vertices $z_i$. Defining a time-order of vertices $v'<v$ if
$z_{v'}^0< z_v^0$ we finally get
\begin{align}
\Gamma(q_1^{\sigma_1},\dots,q_E^{\sigma_E}) 
&= \lim_{\epsilon\to 0}\frac{g^V}{(4!)^V}\!
\int\! \prod_{i=1}^I \frac{d^3 k_i}{(2\pi)^3 2\omega_{k_i}}
\prod_{v=1}^{V-1} \frac{\mathrm{i}
(2\pi)^3 \delta^3 \Big(\sum_{i=1}^I J_{vi} \vec{k}_i 
+\sum_{e=1}^E J_{ve}\sigma_e \vec{q}_e\Big)
}{
\sum_{v'\leq v} \!\Big( \sum_{i=1}^I J_{v'i} \omega_{k_i} 
+ \sum_{e=1}^E J_{v'e} \omega_{q_e}\Big)+\mathrm{i}\epsilon}
\nonumber
\\*
& \times 
\exp\Big(\mathrm{i} \theta^{\mu\nu} \Big(
\sum_{i,j=1}^I I_{ij} k_{i,\mu}^+ k_{j,\nu}^+ 
+ \! \sum_{i=1}^I \sum_{e=1}^E I_{ie} 
\sigma_e k_{i,\mu}^+ q_{e,\nu}^{\sigma_e} 
+ \! \sum_{e,f=1}^E I_{ef} \sigma_e \sigma_f q_{e,\mu}^{\sigma_e}
q_{f,\nu}^{\sigma_f}\Big)\Big)\,.
\label{GammaE}
\end{align}
The vertex which is missing in the product over $v$ is the latest one.
There remain $I{-}V{+}1{=}L$ momentum integrations to perform, where
$L$ is the number of loops.  The integral (\ref{GammaE}) corresponds
to a particular graph with $E$ external and $V$ internal vertices
which all have different dates. The internal vertices are composed of
four different points according to the four fields building the
vertex, with the time-interval within a vertex smaller than the
time-distance to the neighboured vertices. Any external vertex is a
single point which is either later or earlier than all points in
internal vertices. A graph is the connection of each two of these
$4V{+}E$ points by a line which is oriented forward in time, such that
at each point we find exactly one end of a line. We assign to this
graph the integral (\ref{GammaE}) according to the incidence matrices,
which also enter in (\ref{rosette}). Finally, one has to sum over all
different graphs.  Note that a given graph does not have any symmetry
because the four points in the vertices have clearly distinguished
dates. The Feynman rule (\ref{GammaE}) is easily generalised to other
than $\phi^4$ theories. Eq.~(\ref{GammaE}) is the analytic expression
of the Feynman rules listed in \cite{Liao:2002xc}, apart from a
disagreement in the symmetry factor.

We now see that the graph we have computed was very special. Because
of $V{=}1$ the denominator in (\ref{GammaE}) was absent so that the
integration over the propagator momentum $k_1$ was identical to the
na\"{\i}ve Feynman graph computation. This remains true for all
tadpole lines $i$, because for them $J_{vi}k^+_i=0$ for all $v$. For
internal lines connecting points in different vertices we need new
techniques to perform the integrations.

\section{Summary}

As a warm-up for the general treatment we have computed the one-loop
two-point function for a $\phi^4$ theory on noncommutative space and
time in the framework of ``interaction point time-ordered perturbation
theory''. The calculation is based on free fields (on the mass shell),
but at the end the loop momenta become general four-momenta. Our final
result (for that graph) agrees with a Feynman graph computation,
provided that \emph{one assigns to the internal line the real part of the
  Feynman propagator}.  This  can be understood as the inclusion
of acausal processes in the $S$-matrix, because IPTO explicitly
violates causality.  One may speculate that the true time-ordering of
the $\star$-product (\ref{GTxy}) will produce the na\"{\i}ve Feynman
rules involving the standard causal Feynman propagator in non-planar
graphs.  This approach was shown to violate unitarity of the
$S$-matrix. We have thus to decide whether we prefer to give up
(micro-) causality or unitarity in noncommutative field
theories\footnote{Assuming space-time noncommutativity to be a
  model of quantum-gravitational background effects ($\theta \sim
  l_{\mathrm{Planck}}^2$), one can view this abandonment of causality in
  the $\star$-product as its breakdown at the Planck scale.}.

Next we have derived the Feynman rules (\ref{GammaE}) for general
Green's functions.  Power-counting tells us that (\ref{GammaE}) is
expected to diverge if there are subgraphs with $E\leq 4$ external
lines. If there are non-planar divergent graphs, it is not possible to
absorb the divergences by local (hence planar) counterterms. One has
therefore to analyse whether the oscillating phases render the
power-counting divergent integral finite. This requires to develop
techniques for the computation of (\ref{GammaE}) in analogy to the
treatment of the Euclidian case in \cite{Chepelev:2000hm}. Of urgent
interest are the evaluations of the two-loop two-point function and
the one-loop four-point function.

\section*{Acknowledgement}

We would like to thank the Erwin Schr\"odinger Institute in Vienna,
where this paper was finished, for hospitality and for support by the
programme ``Noncommutative geometry and quantum field theory''.

\begin{appendix}

\section{Conventions for Fock space and propagators}

To fix our notation and for convenience we list our conventions for 
free fields and propagators $D^{\pm}(x-y)$ and
$\Delta_F(x-y)$.

The free fields (solutions of the homogeneous Klein-Gordon equation)
are mode-decomposed into negative ($\phi^+$) and positive ($\phi^-$)
frequency parts $\phi(x)=\phi^+(x)+\phi^-(x)$,
\begin{align}
\phi^-(x) &= \frac{1}{(2 \pi)^{3/2}} \int \frac{d^3k}{\sqrt{2\omega_k}} \,
                \mathrm{a}^-_k
                \mathrm{e}^{-\mathrm{i}x_{\mu}k^{+\mu}}\;, & 
\phi^+(x) &= \frac{1}{(2 \pi)^{3/2}} \int \frac{d^3k}{\sqrt{2\omega_k}} \,
                \mathrm{a}^+_k \mathrm{e}^{+\mathrm{i}x_{\mu}k^{+\mu}}\;,  
\end{align}
with the ladder operators $\mathrm{a}^-,\mathrm{a}^+$ obeying 
\begin{align}
  \mathrm{a}^-_k | 0 \rangle = 0 \;, & & \langle 0 | \mathrm{a}^+_k = 0 \;, 
  & & [ \mathrm{a}^-_p , \mathrm{a}^+_q ]= \delta^3(\vec{p}-\vec{q})\;. 
\end{align}
With these definitions we obtain for the two-point vacuum
expectation values and the commutators of positive and negative
frequency parts
\begin{align}
\langle 0 | \phi(x) \phi(y) |0 \rangle = [\phi^-(x) , \phi^+(y)] 
&= D^+(x-y) = \int \frac{d^3k}{(2 \pi)^3 2 \omega_k} \,
\mathrm{e}^{- \mathrm{i}(x-y)_{\mu}k^{+\mu}}\;,
\nonumber
\\
\langle 0 | \phi(y) \phi(x) |0 \rangle = -[\phi^+(x) , \phi^-(y)] 
&= D^-(x-y) = \int \frac{d^3k}{(2 \pi)^3 2 \omega_k} \,
\mathrm{e}^{\mathrm{i}(x-y)_{\mu}k^{+\mu}}\;,
\end{align}
 where $ \omega_k = \sqrt{\vec{k}^2+m^2}$ and 
$(k^\pm)^\mu = (\pm \omega_k,\vec{k})^{\mu}$.
For the Feynman propagator we hence find
\begin{align}
 \langle 0 | \mathrm{T} \big( \phi(x) \phi(y) \big) |0 \rangle 
= \Delta_{F}(x-y) = 
 \int \frac{d^4 k}{(2 \pi)^4}\, \frac{\mathrm{i} 
\mathrm{e}^{-\mathrm{i}(x-y)k}}{k^2-m^2+\mathrm{i} \varepsilon} \;,
\end{align}
and for its complex conjugate 
\begin{align}
\langle 0 | \tau(y^0-x^0) \phi(x) \phi(y) 
+ \tau(x^0-y^0) \phi(y) \phi(x) |0 \rangle = \Delta_{F}^*(x-y) = 
 \int \frac{d^4 k}{(2 \pi)^4}\, 
\frac{-\mathrm{i} \mathrm{e}^{-\mathrm{i}(x-y)k}}{
k^2-m^2-\mathrm{i} \varepsilon} \;.
\end{align}
These propagators are solutions of the
homogeneous and inhomogeneous wave equation, respectively:
\begin{align}
(\partial_\mu\partial^\mu -m^2)_x  D^{\pm}(x-y) &= 0\;, &
(\partial_\mu\partial^\mu -m^2)_x  \Delta_{F}(x-y) &
= -\mathrm{i} \delta^4(x-y)\;.
\end{align}

\end{appendix}

\end{document}